\documentclass[prb,aps,superscriptaddress,endfloats, showpacs]{revtex4}

\begin {document}

\title { Electromagnetic  origin of mass due to folded pseudo-coordinates}

\author {I.E. Bulyzhenkov}

\affiliation {P.~N.~Lebedev Physical Institute, Leninsky pros. 53, Moscow 119991, Russia}
%{}

%\email {ibw@sci.lebedev.ru}

\bigskip
%\noindent {\small {\bf Abstract}
\begin {abstract}
Nullification of the Einstein tensor curvature for the elementary material space with active gravitational field (radial source)  and passive field distribution of its inertial particle (radial sink)  maintains the conceptual equivalence of collinear active and passive energy-momentum densities of the nonlocal energy-matter carrier in General Relativity (GR).  The Ricci scalar curvature in the Lagrangian for electron's material space is related  to  active and passive real mass densities with residual positive energy, while electromagnetic (EM) active and passive imaginary mass densities produce a traceless energy density. 
One four-potential in geometrical sink-source identities universally forms gravitational and electromagnetic fields of the complex mass current.
 EM vector equalities balance anti-collinear four flows of active and passive imaginary masses (called electric charges). Similar GR vector equalities take into account twin coupling of inseparable unlike mass densities existing due to the forth-coordinate symmetry, $ dt = |\pm idx^4/c|$, in the Minkowski four-interval. Two equal pseudo-coordinates $\pm ix^o\equiv \pm x^4 $ serve for equal unlike masses of the same inertial body that results in collinear active and passive  four-currents of real masses with net energy-momentum for collisions and measurements. 
 Couples of unlike twin masses can simultaneously radiate-absorb paired vector waves  with zero-balanced Poynting flows for real-time observations in the two-body gravitational system. Such paired vector waves form tensor gravitons which cannot be screened by third bodies or intercepted by interferometers.
 \end {abstract}
 \bigskip

\pacs{04.20.Cv; 03.50.-z.}
\maketitle

\section {Geometrization of the nonlocal carrier of mass-energy}

The Standard Model has not yet explained the origin of inertial mass and has failed to incorporate Einstein's General Relativity into the $SU(3)\times SU(2)\times U(1)$ gauge scheme. Breaking of Lagrangian chiral symmetry  has been already justified by Quantum Chromodynamics for $10^{-15}$ m scales and it is not clear what further measurements of space beyond the current $10^{-18} m$ limit might change for this mass forming symmetry. The goal of this article is to revisit algebraic and geometrical conditions for the residual creation of the Ricci curvature of GR energy spaces from the traceless electromagnetic (EM) vacuum\cite {Rai} of $U(1)$ gauge fields. Our algebra equally employs both imaginary coordinate branches, $\pm ix^4 \equiv \pm x^o$, of folded  pseudo-geometry for   physical reality with single world time $dt = |\pm dx^o|/c > 0$. Such  4D pseudo-geometry for real folded masses might originated after a spontaneous coordinate folding of regular 4D geometry for  imaginary electromagnetic (iEM) masses with zero-balanced four-flows. Two mirror pseudotimes, $\pm x^o/c$, for the GR four-interval can describe  inseparable spatial coupling of unlike mass densities, $\rho^+= - \rho^-$,  both caring equal and positive mechanical energies of  the {\it gesamt} (= whole)  carrier in single time practice.  The global Big Bang fluctuation of vacuum 4D geometry into the real world pseudo-geometry with one partially folded coordinate (or Big Bang Folding with $[-x^4_{BB}; +x^4_{BB}] \rightarrow 2[0, |ix^4_{BB}|] = 2[0, cT_{BB}]$) might created collinear (residual) energy-momentums and angular momentums vectors from anti-collinear (zero-balanced source and sink)  iEM mass flows. A steady or accelerated relaxation of such a positive mechanical energy fluctuation  to unfolded 4D geometry for electromagnetic vacuum can drive world dynamics of decreasing mechanical masses due to the global decrease of the iEM vacuum fluctuation length $cT \equiv |\pm x^o| \rightarrow 0$. Relaxation shrinking,  $cdT \equiv d|\pm x^o| < 0$, of two folded pseudotimes $\pm x^o$ defines the positive time arrow $dt = |\pm dx^o/c| = dT_U > 0$  for the increasing Universe age $T_U = T_{BB} - T > 0$.  Einstein's tensor gravitation for coupled unlike masses with folded time symmetry (FTS) for two equal pseudotimes $\pm x^o/c$ corresponds to paired vector interactions and can be unified with Maxwell's vector electrodynamics for {\it gesamt} carriers of active and passive complex masses.

Before the 1897 discovery of the electron, Thompson\cite{Tho} proposed to relate the conventional electric field energy $ {\cal E} \equiv ({1/8\pi})\int_a^\infty (-e_o/r^2)^2 4\pi r^2 dr$ $  = e_o^2/2 a $ of the charged sphere to
 corpuscle's mass (or electron's rest-mass $m_o = 0.51$ GeV at $2a$ = $2.8 \times 10^{-15}$ m). He later decided to call the celebrated term $4{\cal E}/ 3$ as the electromagnetic  part of the inertial mass (c=1 unless otherwise stated). However, the Lorentz transformations of the symmetrical EM energy-momentum tensor, $T^{ox} = T^{xo} = [T'^{xo}(1+v^2) + vT'^{oo} + v T'^{xx} ]/(1-v^2) $ $\approx  (4T'^{oo}/3 ) v$,  cannot coincide with the transformations of four-vectors for moving scalar masses. This blocked coherent theoretical attempts to reconcile Thompson's electromagnetic mass with the Lorentz invariant dynamics of the inertial mass $m_o$.
Moreover, there are no reports that electrons exhibited radial inhomogeneities up to the achieved $10^{-18} m $ level for space measurements. Therefore, both theory and practice deny the Thompson radius for the EM content of inertia.
From where does the inertial mass-energy $m_o \neq 0$ come from  if not EM fields or their invariant scalar constructions?

We support Thompson's fundamental guess about the electromagnetic origin of mechanical masses through Sakharov's hypothesis of gravitation as a residual electromagnetic phenomenon\cite{Sak}. For a vector EM-type  unification of two fundamental forces, a joint (forming-up) four-potential for gravitational, $W_\mu$, and electromagnetic, $A_\mu = const \times  W_\mu$, classical fields should be justified. We start this program from the Einstein-Grossmann `Entwurf' flatspace generalization\cite {Ein} of the Minkowski interval and employ their approach  to geometrization of  the gravitational (active) field for similar geometrization of  the continuous particle, which is the $r^{-4}$ radial formation of distributed matter or passive energy. We find  that one  universal four-potential $W_\mu$ can form  Maxwell electromagnetic fields and Einstein-Grossmann metric fields when the balanced Einstein curvature, $G_o^o \equiv g^{o\nu}R_{o\nu} - R/2 = 0$ for static radial mass-energy, describes joint geometrization of the elementary continuous source (active mass-energy which attracts other particles) and the elementary continuous sink (passive mass-energy which is attracted by other sources). Then the Ricci density  $R = g^{\mu\nu}R_{\mu\nu}$ describes  sum of equal active and passive scalar mass densities of the continuous material carrier of distributed mass-energies. The local metric tensor $g^{\mu\nu}$ of pseudo-Riemannian 4D geometry with a flat 3D sub-interval within the $r^{-4}$ nonlocal carrier of distributed elementary energy depends on its integral value called the GR energy-charge \cite{Bul}.

The energy-momentum  $P_\mu \equiv m_p u_\mu \equiv m_p g_{\mu\nu} dx^\nu /ds = K_\mu + G_\mu P_o $ of a non-rotating
probe particle with passive-inertial mass $m_p$ and the positive kinetic energy $K \equiv m_p / {\sqrt {1-v^2}} $ complies with the four-component gravitational potential, $G_\mu \equiv (G_o, G_i)$, for the probe (passive-inertial) charge-energy,   $P_o/c^2 = m_p {\sqrt {g_{oo}}}/ {\sqrt {1-v_iv^i}} > 0, $  in pseudo-Riemannian space-time with strict spatial flatness, where \cite{Bul}
\begin {equation}
 ds^2 \equiv  g_{\mu\nu} dx^\mu dx^\nu  = \left ( \frac {dx^o + G_i dx^i }{1-G_o} \right )^2
 -\gamma_{ij} dx^i dx^j,
\end {equation}
$g^{oo} = (1-G_o)^2 - \delta^{ij} G_iG_j$, $g^{oi}= g^{io} = \delta^{ij}G_j $, $g^{ij} = - \delta^{ij}, g_{oo} = (1-G_o)^{-2}, g_{oi} =  G_i/(1-G_o)^2, g_{ij} = - \delta_{ij} + G_iG_j(1-G_o)^{-2}$, $v_i = \gamma_{ij}dx^j/d\tau$, $d\tau^2 = ds^2 + \gamma_{ij} dx^i dx^j$,  $\gamma_{ij} \equiv g_{oi}g_{oj} g^{-1}_{oo} - g_{ij} =  \delta_{ij}$. Such flatspace geometrization of the $r^{-4}$ continuous particle together with geometrization of its $r^{-2}$ Newtonian field `simplifies' ten empty-space Einstein equations for point masses up to four non-empty space equations for moving mass-energy densities \cite {Bul, Buly}:
\begin {eqnarray}
{c \over \kappa} \left( {2 g^{\mu\nu}R_{o\nu} } - {R u_o u^\mu }\right )  = 0 
\end {eqnarray}

In this energy-to-energy attraction of radial mass-energy carriers, the Newtonian inverse square force $  ({\hat {\bf r} G}/r^2)(-P_o^a/ c^2) (P_o^p/c^2)$, exerted to the positive passive-inertial charge, $P^p_o/c^2 > 0$, takes place due its attraction by the active (negative) gravitational charge $ (-P^a_o/c^2) < 0  $, which also possesses the positive active self-energy $(-P^a_o/c^2) (-c^2) = P^a_o > 0$.
Our ultimate goal is to explain the origin of negative and positive GR charges with collinear energy-momentums in terms of residual parts of  anti-collinear (balanced) iEM mass four-flows. We tend to develop pure geometrical approach to electron's active and passive complex masses with imaginary electromagnetic counter-flows (balanced densities) and GR real co-flows (residual densities). Below we discuss the residual EM origin of the electron's GR four-momentum and the Ricci curvature of electron's non-empty space with the dynamical relations   (2) between gravitational and inertial (mechanical) energy densities. Notice in (2) that $(Ru_o/2\kappa) cu^\mu$ is the GR energy flow density of the scalar active+passive mass densities, while $(R/2\kappa)cu^\mu$ is four-current of these mass densities. Therefore, despite all four-components of the  gravitational/inertial fields in (2) correspond to four-flows of energy, this gravitational equation can be submitted for mass four-currents,
$\rho u^\mu = R^\mu_o / \kappa u^o$, like in vector electrodynamics.

 For static fields $G_o = - r_o/r$ and $G_i = 0$, the radial source generates a negative interaction potential\cite{Bul} $W_o(r) = - c^2 ln ( 1 + r_o r^{-1} )$ due to its active (negative) gravitational charge
$ m_a {\sqrt {g_{oo}}} /{\sqrt {1 -v_iv^i}} \equiv P_o/(- c^2) < 0$ with the negative mass scalar, $m_a = - m_p < 0$.
 Such a gravitational charge with the negative active mass can attract positive passive-inertial charges through vector-type interactions only due to $\pm mass - \pm x^o$ pseudotime  symmetry, which is hidden in single time reality for only positive mass-energies.  In other words, our goal is to find justifications for new physical options when two equal pseudo-coordinates, $\pm x^o$, and two unlike masses keep $MPX^o$ invariant vector forces. This might be possible when parity P and unlike masses M are associated with different pseudo-coordinates $\pm x^o$, rather than with one real time. Such FTS relations $m^+dx^o = m^-d(-x^o)$  for twin coupled unlike masses (comasses, for brief) keeps the positive time rate $dt = |\pm dx^o|$ under the inversion $x^o \rightarrow - x^o $ which does not reverse the real world motion of observed {\it gesamt} carriers of energy.
   The static radial potential $W_o(r)$ with the rest-mass driven scale $r_o \equiv G |m^{\pm}| > 0$ can describe both a radial astrodensity, $\rho_a = -\nabla_i^2 W_o(r)/4\pi G c^2 < 0$ of the active gravitational charge $P_o/(- c^2) < 0$, and the passive-inertial mass-energy density, $\rho_p = (\nabla_i W_o)^2/4\pi G c^2 > 0$, of the passive gravitational charge $P_o/c^2 > 0$. Similarly to electrodynamics, negative active GR charge density $\rho_a$ gains positive self-energy $\rho_a (-c^2) = \mu_a > 0$ in its negative self-potential $(-c^2) = -1$.  Constant self-potentials $\mp c^2$ for active/passive GR charges are free from self-forces because $\nabla (\mp c^2) \equiv 0$.
Active and passive rest-mass densities equally contribute\cite {Bul, Buly}  to the local Ricci scalar  $Rc^2 =\kappa [\rho_a  (-c^2) + \rho_p (+c^2) ] =  \kappa ( \mu_a + \mu_p) = 2\kappa \mu_a = 2 \kappa \mu_p > 0$, where $\kappa \equiv 8\pi G/c^2 = 1.86\times 10^{-26} m/kg $. Our goal is to find how the Ricci curvature R of flat material space, filled by radial active and passive masses, could match complex active and passive mass flows with zero contributions of imaginary masses into inertia and gravitation. If so, electric charges may be associated with imaginary masses in the FTS unification of classical GR+iEM fields of complex masses (or complex charges with residual real self-energies).

  All densities of the unified carrier of real energy with complex GR+iEM masses should obey a universal transformation law, otherwise electric charges and mechanical masses would split under 3D translations. Therefore, four relativistic balances  (2) between active/passive fields and four-currents of gravitational/inertial energy densities request for the charged radial carrier similar balances between electromagnetic fields and energy four-currents of electric charge densities. In what follows, we assign   passive (sink) self-energy to continuous imaginary mass or electric charge  and active (source) self-energy to iEM field, with an exact local balance of sink and source counter-flows for their inertia-free energies.  Basic geometrical identities for radial fields with the universal four-potential $W_\mu$  will be derived for the unified carrier with radial GR+iEM masses. These vector identities will equally work for real (gravimechanical) and imaginary (electrical) parts of elementary mass currents if one employs  both pseudo-time coordinates $\pm x^o$ for description of single time reality. The identities for imaginary masses can reproduce Maxwell's equations-equalities and for real masses can reproduce   the  Newton-Poisson's gravitation equation due to folded time symmetry. Zero divergence of complex mass four-flows under arbitrary fluctuations and wave modulations of the nonlocal (radial) particle can shed some light on the particle-wave duality of matter in theory of classical fields.

    By balancing anticollinear active and passive iEM four-flows of the distributed (continuous source-field  and continuous sink-field) radial electron, we should reserve physical options for residual four-flows of its real mass density. The latter should maintain a collinear transport of positive energy, $(-P_o/c^2)(-c^2) = (-\int dr 4\pi r^2 (\nabla_i^2 W_o/4\pi G c^2)(-c^2)$ $  > 0 $,  of the active (negative) GR charge and equal positive energy of the passive-inertial (positive) GR charge, $P_o/c^2 =  \int 4\pi r^2 dr  (\nabla_i W_o)^2/4\pi G c^2  > 0$. Residual or collinear active (source) and passive (sink) energy-momentums in physical reality require FTS partners in the world pseudo-geometry for balanced (zero sum) creation of real energy from void nothingness. Such a folded time partnership can explain the creation of  active masses with positive GR energies simultaneously with equal positive energies of passive masses. This FTS partnership assumes single time reality of radial mass-energy densities due to  pseudo-geometrical folding of 4D vacuum geometry of traceless EM energy spaces.
   
Again, one should assign mirror $\pm x^o$ branches in folded pseudo-geometry to inseparably overlapping unlike real masses of the radial electron. These twin masses (comasses $m^{\pm}$) get identical (folded) time dynamics and attractive cross-branch forces in vector interactions with  $MPX^o$ symmetry. This folded time symmetry suggests to read the GR mass-vs-energy relation $P_\mu P^\mu = (\pm m_o)^2(\pm c^2)^2$ equally with both algebraic signs next to the inertial mass $m_o$, with equal gravitational $(-m_o)(-c^2)$, and inertial, $m_o c^2$, rest-frame energies. What benefits can  the cross-branch symmetry $m^+dx^o \equiv m^-(-dx^o)$ provide for the mass creation mechanism of inertia-free electric charges?   Geometrical folding of the zero-sum vacuum 4D fields with anti-collinear four-momentums enables cross-branch pairing of twin comasses  with  non-zero net energy. This residual energy is positive  under the positive time arrow. Briefly, a short fold of zero-sum anti-collinear vectors (in void balance of iEM vacuum states or imaginary mass counter-flows) results in a residual four-vector (observed energy-momentum in pseudo-metric reality) next to the iEM vacuum parts of charged matter. Interactions of two imaginary masses (electric charges) result in real forces exerted on real (inertial) mass-energies. Real EM forces and real EM energies can be observed exclusively through dynamics of real GR masses. Electric charges without inertial rest-masses (or pure imaginary masses) were never observed in practice.

\section {Locally balanced active and passive imaginary mass flows}

  There is no satisfactory mathematical approach to the diverging field energy of the Coulomb point charge in theory of classical fields.
Therefore, we consider electron`s classical charge as a continuous (field) distribution with passive electric energy, while electron`s classical field as a charged continuous distribution with equal amounts of active electric energy. And we assign imaginary electric densities to both the passive (sink) radial charge-field and the active (source) radial charge-field that keeps real interaction forces between active, $(iq)^*$, and passive, $iq$, electric charges. In order to be more specific,  we replace the operator particle density
  $\delta({\bf r}- {\bf R}_c)$ with the static \cite {Bul}  particle density $n({\bf r}- {\bf R}_c) = r_o /4\pi({\bf r}- {\bf R}_c)^2 ( |{\bf r}- {\bf R}_c| + r_o )^2 $ around the center of spherical symmetry at ${\bf R}_c$. In analogy with the constant self-potential $c^2$ for real mass densities, the radial density of  imaginary mass or electric charge, $i^*\rho(r) = i^*(-e_o)r_o/4\pi r^2 (r + r_o)^2 = i^*\nabla {{\bf d}}/ 4\pi $, also has constant  self-potential, $i\Lambda_e = const$, for generation of real EM self-energy in all material points. Then, the local balance of active (source) and passive (sink) real EM energy densities can be achieved for the radial (astro)electron in full analogy with the Equivalence Principle for active and passive GR mass-energy densities \cite {Bul, Buly}.

   Thompson's electric field energy  does not contribute to collinear gravitational/inertial energy-momentums  of  the radial electron because this self-energy of active electric fields is locally balanced by self-energy of the passive radial charge. This strict local balance within the continuous electron bypass the energy-divergence problem of the point electron. Therefore, one can rigorously consider electron`s densities at extremely small spatial scales, including the half-charge scale $r_o =  Gm_o (= 7\times 10^{-58}m $) for the imaginary electric density $i\rho (r)$, which generates the post-Coulomb interaction potential, $i^*A_o(r)=$ $i^*(e_o/r_o)W_o(r)$ = $i^*(-e_o/r_o)ln(1+ r_or^{-1})$, for attraction/repulsion of other electric charges. This inhomogeneous imaginary potential for real interactions with other passive imaginary charges differs from the uniform self-potential $i^*\Lambda_o$. The latter  nullifies local self-forces ($i^*\nabla \Lambda_o \equiv 0$) without needs in additional   Poincar{\'e} pressures for stabilization of the elementary charge density $i\rho (r)$.  The constant self-potential $i^*\Lambda_o$ transforms, in fact, the passive elementary charge density $i\rho(r)$ into the real self-energy density $i\rho (r)\times i^*\Lambda_o $ = $\rho (r) \Lambda_o = {\bf d}^2(r)/4\pi $. The self-potential magnitude, $\Lambda_o \equiv (-e_o)/r_o = - 4\times 10^{42} V$, for the radial (astro)electron can be defined from the Maxwell static field solution $ id_r =  - i \partial_r A_o(r) =   i(-e_o)/r(r + r_o) $ due to the balanced active and passive electric field energies, with  $ (-e_o)  \Lambda_o = \int dr 4\pi r^2 \rho(r) \Lambda_o = \int dr 4\pi r^2 d^2_r/4\pi =  \int dr 4\pi r^2 \rho(r) A_o(r)  =4\times 10^{42} eV $.

 Once static post-Newton and post-Coulomb  potentials of active GR and EM charges are proportional to each other, $W_o / A_o =  Gm_o/e_o = - c^2/\Lambda_o = const$, than one can use identical covariant transformations for gravitational and electromagnetic four-potentials, $W_\mu(-\xi^2)=-u_\mu ln [1 + (r_o/ {\sqrt {-\xi^2}} )]$ and $i^*A_\mu(-\xi^2) $=$i^*(e_o/r_o) W_\mu(-\xi^2)$ $ = - i^*(e_ou_\mu/r_o)ln [1 + (r_o/ {\sqrt {-\xi^2}} )]$, respectively, with $-\xi^2 \equiv - \xi_\mu\xi^\mu = -x_\mu x^\mu + (u_\mu x^\mu)^2 \Rightarrow  r^2$ for $u^\mu = (1,0,0,0)$, $\xi^\mu = x^\mu - u^\mu(u_\nu x^\nu)$, $u_\mu u^\mu = 1 $, $u_\mu \xi^\mu  = 0 $, $\nabla^\mu (-\xi^2)/2 = - x^\mu + u^\mu (u_\nu x^\nu)$.  

The Hilbert gravitational action $S = - \int {\sqrt {-g}} d^3x dt  R/ 2\kappa $  for the geometrized (astro)carrier in non-empty space depends on the active (continuous source) and passive (continuous sink)  contributions into the Ricci scalar density  $R = \kappa \mu_a + \kappa \mu_p > 0$ in curved space-time under strict spatial flatness. Such an active+passive Ricci structure  may  originate, in principle, as a residual real part of complex active and passive masses in the unified GR+iEM
action $S_{_Z} = \int  {\sqrt {- g}} d^3x  dx^o (i^*Z^*+iZ) /2\kappa $ for distributions of active, $i^*Z^*$,  and passive, $iZ$, complex mass densities of the elementary continuous carrier of {\it gesamt} energy.
We associate the active complex charge distribution with the continuous source (including its  wave modulations) and the passive complex charge distribution with the continuous sink in the Lagrangian.
In this way, we can generalize the real Ricci scalar, $R /\kappa = \mu_a  + \mu_p = -(i^*Z^* + iZ)/\kappa $, in the Hilbert action on the sum of complex active $(iZ)^*$ and passive $iZ$ Lagrange densities in coordinate 4D space,
\begin {equation}
S_{_Z}=    \int_\infty^\infty\!{d^3x}
 \left [ \int_0^T {\sqrt {- g}} idx^o \frac {(-Z^*+Z)}{2\kappa} + \int^0_{-T} {\sqrt {- g}} idx^o \frac {(-Z^*+Z)}{2\kappa} \right ]
\equiv S^+_{_Z} + S^-_{_Z}\end {equation}

Now we can set the complex scalar density $i^*Z^* = i^*e^{i\gamma}f_{\mu\nu}f^{\mu\nu} M/m_o $ for the active (source) fraction of elementary matter and $iZ = ie^{-i\gamma}f_{\mu\nu}f^{\mu\nu}M/m_o $ for its passive (sink) fraction.
Hereinafter anti-symmetrical tensor intensities, $f_{\mu\nu} \equiv (\nabla_\mu W_\nu $  -  $\nabla_\mu W_\nu)_{\mu\neq \nu}$ and $f^{\mu\nu} \equiv  g^{\mu\rho} g^{\mu\lambda}f_{\rho\lambda}$, and the contravariant potential $W^\mu = g^{\mu\nu}W_\nu$ are defined through the dimensionless four-potential $W_\mu$.
The real parameter $M$ is an arbitrary magnitude of the complex GR+iEM mass composition, while $Msin \gamma = m_o = 8\pi r_o/\kappa$ is its real GR (gravimechanical) part.
%part

At first we drop for a moment the `unphysical' negative branch $x^o < 0$ in the elementary action (3) for complex mass densities in 4D space. The reduced action with only one positive branch for such transition in question to 3D+1T pseudo-geometry with unfolded physical time, when $dt = + dx^o/c$ (rather than $dt = \pm dx^o/c$) and
\begin {equation}
S^+_{_Z}=
   - \int_\infty^\infty\!  {d^3x} \int_0^{T}  {i{\sqrt {-g}} dt }   \frac  {M}  {16 \pi r_o}  \left (
  e^{+i\gamma} {f_{\mu\nu} f^{\mu\nu}}
   -   e^{-i\gamma}{f_{\mu\nu} f^{\mu\nu}}   \right ),
\end {equation}
employs the incomplete Lagrangian without FTS options for the negative (real) mass.
Equal contributions to the active matter density (with $e^{+i\gamma}$) and the passive matter density  (with $e^{-i\gamma}$) should not be subtracted in (4) before the Lagrange variations for dynamical equations. Evaluations of covariant derivatives, $\nabla_\nu W^\mu = -r_o \xi_\nu u^\mu /({\sqrt {-\xi^2}} + r_o ) (-\xi^2)$, can be achieved under the Lorentz condition $\nabla_\mu W^\mu = 0$. Notice that  $\nabla^\mu W^\mu \neq 0$  or  $\xi^\mu u^\mu \neq 0 $ even for  $\xi_\mu W^\mu = \xi_\mu u^\mu = 0$.

The small dimensionless factor $\gamma $ will be related to the residual EM mechanism for  gravitation/inertia of the electron ($\gamma\ll 1$ and $ m_o \ll M$. This factor $\gamma \equiv \gamma (T)\approx const $ imbalances complex mass densities of active    ($iMe^{i\gamma}$)
and passive   ($iMe^{-i\gamma}$) fractions of the elementary continuous carrier. One can associate equal active and passive imaginary parameters
       $Im (iMe^{i\gamma} ) = M cos \gamma$  and
$Im (iMe^{-i\gamma} ) = M cos (-\gamma)$, respectively, with active and passive EM masses of an  electrically charged carrier. Real parts,  $Re (iMe^{i\gamma} )\equiv m_a $ and $Re (iMe^{-i\gamma}) \equiv  m_p$, can be associated with active (source) and passive (sink) GR masses, respectively, with
$(-m_a) = m_p = M sin\gamma = m_o > 0$.
 Variations of the action (4) with respect to the four-potential $W_\mu$ ought to produce Euler-Lagrange equations for imaginary and real mass four-flows. Indeed, Lagrange equations for the imaginary EM mass densities from the action (4)  maintain the mutual local balance of anti-collinear active and passive EM four-flows in terms of vector identities for equal active and passive imaginary masses or electric charges , $i^*q_a \equiv i^*M cos \gamma  {\sqrt G}$ and  $iq_p \equiv iM cos (-\gamma ) {\sqrt G}$, respectively,
\begin {eqnarray}
%\cases {
 \frac {(-iq_a)^* \nabla_\nu f^{\mu\nu}} { 4\pi r_o}  \equiv
 \frac {{iq_p}\nabla_\nu f^{\mu\nu} }{4\pi r_o}
  \equiv
   {iq_p }  n u^\mu \equiv
  {(-iq_a)^* }  n u^\mu,
 %}
 %   \cr\cr
%4\pi (- sin \theta e_o) n u^\mu =   - (- sin \theta e_o) \nabla_\nu f^{\nu\mu} .\cr
\end {eqnarray}
where $q_a = q_p = q_o $ is a real number and $ q^2_o = GM^2  - Gm_o^2$.

    Here we first equalized  the covariant derivatives for local counter-flows of active and passive imaginary mass (or iEM charge) densities. Then we set the passive  four-current through the four-flow $nu^\mu$ of the radial particle density $n$.
    The passive four-current locally balances (reiterates) the active one.
    One may use the local Lorentz condition  $\nabla_\mu A^\mu =  const \nabla_\mu W^\mu = 0$ for steady radial states with $r_o = const$ in the unified iEM and GR four-potential $W^\mu$. The particle density flow
    $nu^\mu $  for $r_o =const$ and $\xi_\mu W^\mu = 0$ of the steady elementary source can be specified through the following evaluations of covariant derivatives,
  \begin {eqnarray}
%\cases
  n({\sqrt {-\xi^2}}  ) u^\mu \equiv \frac{1}{4\pi r_o} \nabla_\nu[{\nabla^\mu W^\nu ({\sqrt {-\xi^2}}} )
 - { \nabla^\nu W^\mu ({\sqrt {-\xi^2}}}  ) ] =
  {\nabla_\nu } \left [ \frac {(\xi^\mu u^\nu - \xi^\nu u^\mu) }{ 4\pi ({\sqrt {-\xi^2}} + r_o )\xi^2}\right ]  \cr = \frac {u^\mu}{ 4\pi ({\sqrt {-\xi^2}} + r_o )\xi^2}   -
 \frac {({\sqrt {-\xi^2}} + 2r_o ) u^\mu}  { 4\pi ({\sqrt {-\xi^2}} + r_o )^2\xi^2}
    = \frac {r_o}{ 4\pi  ({\sqrt {-\xi^2}} + r_o )^2 (-\xi^2)  } u^\mu.\cr
    \end{eqnarray}
       The particle's density $n(-\xi^2)$ matches the electric charge density  $(-e_o)n(r) = (-e_o)r_o/4\pi (r+r_o)^2r^2$  and the inertial mass-energy density $\mu_p (r) = m_o c^2 n(r)$ in the electron's rest frame\cite {Bul}, where ${\sqrt {-\xi^2}} = r$. The post-Coulomb electric field intensity, $(e_o/r_o) f^{oi}=e_ou^o x^i /(r+r_o)r^2= d^i$, is coherently derived from $f^{\mu\nu} \equiv  {(\xi^\mu u^\nu - \xi^\nu u^\mu )r_o } /{ ({\sqrt {-\xi^2}} + r_o )\xi^2}$ in this static limit with $u^o = 1$ and $4\pi n(r) e_o = \nabla_i d^i$. The passive imaginary counter-flow is identical to the active flow in (5) and, therefore, the left-hand side covariant derivative in (5) for active tensor fields can ultimately be  interpreted thorough the  source flow $n u^\mu$, like the passive  continuous particle through the similar siink flow.  Passive and active  four-flows  are bound identical entities in this approach, while their tensor fields can be shifted,  $f_p^{\mu\nu} -  f_a^{\mu\nu} = {\tilde f}^{\mu\nu} \neq 0  $, by non-trivial  (wave) solutions of  $\nabla_\nu {\tilde f}^{\mu\nu} = 0$.

  The continuous source and sink structures may differ from the radial particle structure $n({\sqrt {-\xi^2}}  )$ due to fluctuations,  
    radial waves  or other  inhomogeneous modulations of $n({\sqrt {-\xi^2}}  )$. Time varying interactions with real energy exchanges or radiation energy losses/gains  for a selected body can result in variations of its radial scale $M_oc^2/G$ and, consequently, in violations of the  Lorentz steady-state condition  $\nabla_\mu W^\mu =  0$. The equations (5) are source-sink equalities for iEM mass four-flows or electric four-currents.  They obey the strong conservation condition $\nabla_\mu (n u^\mu) \equiv 0$ for fluctuating matter of continuous sources and sinks, which can exhibit particle-wave duality of  elementary energy distributions.

 %cor
  The four-flow equalities (5) for active and passive field distributions were derived for  imaginary (EM) radial masses or radial electric charges. These  non-empty space field equalities  can  be deduced to empty-space Maxwell equations for point charges after the formal (operator) simplification of the elementary radial density $n (r) \Rightarrow \delta (r) $. 
 The strict equivalence of active and passive electric charges (or imaginary masses) of one elementary particle in the  Lagrange equations for the iEM densities (5) yields  the following  vector, tensor, and scalar identities,
            \begin {eqnarray}
      \cases {
      n (-\xi^2)u^\mu(x)   \equiv \nabla_\nu f^{\mu\nu}(x) /4\pi r_o \cr
       \nabla_\lambda f_{\mu\nu}(x) + \nabla_\nu f_{\lambda\mu}(x) + \nabla_\mu f_{\nu\lambda}(x) \equiv 0 \cr
      \nabla_\mu [n (-\xi^2) u^\mu(x)]  \equiv 0,
         \cr
      }        \end {eqnarray}
      for the forming-up  dimensionless densities of active/passive fractions of the elementary continuous carrier of real self-energies. These elementary geometrical structures should be properly loaded  with active/passive iEM and GR charges.

    Any wave modulation of the radial electron should be considered within its infinite material space.   The geometrical flow conservation, $ \nabla_\mu (nu^\mu) \equiv 0$, under mutual transformations  of radial and wave material structures can address the particle-wave duality of the continuous electron.  Such non-empty space physics of nonlocal elementary matter-energy admits distant timeless influences of overlapping continuous particles. There are no free space regions according to the local dimensionless identities (7) for nonlocal elementary carriers of energy. Therefore, classical field waves are, in fact, material space  modulations. They can be associated with one or another rest-mass carrier of energy and with timeless switching between nonlocal radial hosts.
             Electrodynamic equations for a statistical ensemble of k overlapping continuous               particles
              and their EM fields are followed  from (7) or (5) as the local superposition of elementary densities of iEM mass flows or electric currents. But how can elementary imaginary vector flows of one radial electron be related to its real mass
             flows?

\section {Folded time symmetry for the residual EM origin of inertia and gravitation}

If an inertial carrier of energies with real passive/active masses $M sin\gamma = m_p = -m_a $ and distributed densities
$m_{(p/a)} r_o/ 4\pi r^2 (r+r_o)^2$ has no a mechanical counter partner, than self-translations or self-rotations of such an infinite material distribution might violate conservation laws or not make much sense in relativistic physics. Mach\cite{Mac}  proposed to consider motion of a passive mechanical mass with respect to `the rest of the Universe' or to nullify inertia without this `rest'. We follow this 1904 eureka for counter-motion of inertial masses in order to justify mirror time dynamics for locally converted real masses in folded pseudo-geometry of the Universe. Twin unlike masses (comasses) of a mechanical body obey identical space translations and rotations or identical events under the real time arrow. Pseudo-coordinate folding of two anti-collinear and  zero-sum imaginary mass-flows can result under real time in finite sum of collinear (folded) four-vectors for mechanical energy-momentum. The Machian-type mechanism  for the positive mass-energy $m_o$ creation (under the  positive time rate $  dt = |\pm idx^4|> 0$) takes place due to a fluctuation fold of forth coordinate $x^4$ in iEM vacuum 4D geometry.
This geometrical fluctuation transfers two balanced imaginary mass counter-currents with traceless (zero-sum) densities into real mass-energy currents in observable reality with pseudo-geometry. Such a pure geometrical creation of non-zero four-vectors for inertial masses still obeys Machian counter-motion with opposite four-momentums and opposite angular momentums in unbroken EM vacuum geometry.
Inertia of every folded geometrical formation (with folded four-momentums and folded angular momentums) is defined by its GR energy-charge $P_o = m_o {\sqrt {g_{oo}}/{\sqrt {1 - u_iu^i } }}$ in flatspace Machian relativism \cite {Bul, Buly}. Therefore, inertia of every mechanical body in a gravitational system depends on spatial distributions of other bodies. Matter `there' determines inertia `here' in the central point of Mach's qualitative analysis. In addition to this point, initial creation of  observable mechanical energies is also related to Mach's quest for mutual relativism of masses which are paired as twin comasses in physical reality due to world pseudo-geometry.

The EM counter-flows (5) of equal active (source) and passive (sink) imaginary  masses ({\it i.e.}electric charges) are balanced. These iEM mass flows would never be observed in physical reality unless imaginary masses gain real residual parts for inertial collisions with real energy-momentum and angular momentum exchanges.  Recall that  electric charges without inertial masses were not found in practice.  Utilization of real EM energies is possible due to their transformation into mechanical energies with further collision exchanges between real rest-mass bodies. The key question is how can the imaginary vacuum mass-flows (5) with identically balanced source and sink four-vectors gain a residual four-momentum imbalance without violation of the mathematical equalities (7) for geometrical flows of active/passive material  structures? 
To answer briefly, FTS of real world pseudo-geometry with the single time arrow,  $ dt = |\pm idx^4| > 0 $ for folded pseudo-coordinates $\pm idx^4$, can provide the Machian-type cross-balance for inseparable coupled unlike masses and their fields without violation of the dimensionless identities (7).
Folded pseudo-coordinated in 4D geometry with directed time stand behind the residual GR mass creation from inertialess iEM masses. Such pseudo-geometry enables observable practice with real EM and GR energies.

Let us return to the point that the imbalance $\gamma$ in the reduced `physical' action (4)  should describe real  parts of active and passive complex masses.  The one branch action (4) exhibits positive Ricci curvature, $R = \kappa (-c^2)\rho_a + c^2\kappa\rho_p = \kappa \mu_a + \kappa \mu_p > 0$, for the positive passive-inertial $\rho_p = \mu_p/c^2> 0$ and negative active, $\rho_a = -\mu_a/c^2 = -\mu_p/c^2  < 0$, mass densities. In other words, the residual Ricci imbalance of complex GR+iEM masses in (4) should result in the negative active mass, $m_a = -m_o < 0$, of the source (active field or the conventional classical field) and the positive passive-inertial mass $m_p = m_o > 0$ of the sink (passive field or the continuous radial particle). Such a carrier of unlike active and passive masses with equal positive energies does possess imbalanced or collinear mass-energy four-flows according to the conceptual source/sink equalities (7).  The only challenge is to explain the origin of the negative active mass next to the positive passive-inertial mass of a real-time body, which is a carrier of elementary distributed energies with equal source and sink fractions.

First we make direct evaluations of the Lagrange density in (4) for the steady radial carrier ($r_o = const, \nabla_\mu W^\mu = 0$) and compare this action for complex GR+iEM masses with the Hilbert gravitation action $S= - \int {\sqrt {-g}}d\Omega R/2\kappa$ for an elementary gravitational field,
\begin {eqnarray}
 -i\frac {{\sqrt {-g}M } (e^{i\gamma} - e^{-i\gamma}) }{16\pi r_o}   f_{\mu\nu}f^{\mu\nu}   = - i\frac { {\sqrt {-g} M}(e^{i\gamma} - e^{-i\gamma}) } {16\pi r_o  }     \frac {r^2_o (\xi_\mu u_\nu - \xi_\nu u_\mu)(\xi^\mu u^\nu - \xi^\nu u^\mu)} { ({\sqrt {-\xi^2}} + r_o )^2 (-\xi^2)^2} \cr\cr
=     i\frac {   (  e^{i\gamma}  - e^{-i\gamma} ){\sqrt {-g}} M n } {2 } \Rightarrow    - \frac {{\sqrt {-g}}R}{2\kappa}.\cr
\end {eqnarray}

From here  ${\sqrt {-g}}R/2\kappa $$ = ( M sin\gamma )  n {\sqrt {-g}}$ or the Ricci scalar, $R = 2\kappa m_o n \equiv \kappa (|m_a|+m_p)n$, is proportional to the summary ({\it gesamt } = whole) density of gravitational (active) and mechanical (passive- inertial) masses of the radial particle. It is important to understand for moving material spaces with flat 3D intervals \cite {Buly} that ${\sqrt {-g}} = |ds/dx^o| \equiv {\sqrt {g_{oo}}} (1 - g_i u^i) {\sqrt {1 - u_iu^i}} $, rather than 
 ${\sqrt {-g}} = {\sqrt {g_{oo}}}$ for the Hilbert action of static field distributions. Therefore, the Lagrangian of the moving GR field results (in the absense of rotations) in the Special Relativity Lagrangian for a moving distribution of the elementary mass $m_o$, ${\sqrt {-g}}R/2\kappa $ $\Rightarrow m_o n {\sqrt {1-u_iu^i}}$.

 Evaluations of the energy-momentum tensor $T_\mu^\nu$ for the complex GR+iEM mass with active source and passive sink may count only residual gravimechanical energy densities,
  \begin {eqnarray}
T^\nu_\mu \equiv \nabla_\mu W_\rho \frac {\partial (i^*Z^* + iZ)}{\partial (\nabla_\nu W_\rho)}  - \delta^\nu_\mu (i^*Z^* + iZ)
=   \frac {  (  e^{i\gamma}  - e^{-i\gamma} )  m_o n} {2 i sin\gamma } \left ( {\delta_\mu^\nu}  - \frac {2\xi_\mu \xi^\nu} {\xi^2}  \right ),
\end {eqnarray}
  {\it i.e.} only GR tensor components with the positive trace $T^\mu_\mu = R/\kappa > 0$.

The Euler-Lagrange formalism for the `physical' time-branch action (4)  cannot result
 in a required variational balance for equal and finite collinear four-momentums of negative active,  $m_a = - m_o$, and positive passive, $m_p =  m_o = - m_a$, masses because of their non-vanishing active and passive mass four-flows,
 \begin {equation}
I^\mu \equiv I^\mu_p + I^\mu_a \equiv   m_p n u^\mu  +  \nabla_\nu
 \left (\frac {(-m_a) f^{\mu\nu}}{4\pi r_o}  \right )\equiv 2m_o  n u^\mu \equiv 2 \nabla_\nu
  \left (\frac { m_of^{\mu\nu}}{4\pi r_o} \right ) \neq  0,
 \end {equation}
due to (7).
Equalities in (10) manifest equal collinear GR densities  for mechanical (inertial) and gravitational energy flows. Lunar Laser Ranging and other gravitational tests could test, in principle, this  equivalence for mechanical and gravitational four-vectors.  According to (10), any variations of the gravitational (active) field energy of the Moon should coincide with variations of its inertial (passive)  energy in Earth's and Sun's fields.

The classical variational formalism for Lagrange dynamics demands (due to the non-vanishing four-current  (10) with $m_a = - m_p$ and $I^\mu_a = I^\mu_p$) to keep the negative pseudo-time branch in the geometrical action (3). Then,  space-time dynamics of complex mass flows with residual real masses can correspond
FTS  Lagrange dynamics under the single time arrow. The positive and negative $x^o$ coordinate branches in (3) provide a new physical opportunity to relate active and passive masses to the opposite coordinate rates, $\pm dx^o$, and to read the GR mass-energy equality $P_\mu P^\mu = (\pm m_o)^2$ with both algebraic signs. Folded time  symmetry of real residual masses is  hidden for real-time observers of vector-type gravitational interactions. Nonetheless, the Euler-Lagrange variational balance from the complete geometrical action (3),
\begin {eqnarray}
I^\mu(+x^o) + I^\mu(-x^o) \equiv \left (m^+_pnu^\mu  +  \nabla_\nu  \frac {(-m^+_a) f^{\mu\nu} }{4\pi r_o} \right )
+ \left ( m_p^-nu^\mu  +  \nabla_\nu  \frac {(-m^-_a) f^{\mu\nu} }{4\pi r_o}\right ) \cr\cr
 \equiv\left[m^+_pnu^\mu  -  \nabla_\nu  \frac {m^-_a f^{\mu\nu} }{4\pi r_o}\right ] +
\left [m_p^-nu^\mu - \nabla_\nu  \frac {m^+_a f^{\mu\nu} }{4\pi r_o} \right ]  \equiv 2(m_p^+ + m_p^-)nu^\mu = 0,
\end  {eqnarray}
analytically reveals  twin unlike active and twin unlike passive masses in one {\it gesamt} carrier of active and passive GR energies.
Here equal cross-branch masses $m^+_a (+x^o) \equiv  m^-_p(-x^o) = - m_o < 0$ and
$  m^-_a (-x^o) \equiv m^+_p(+x^o) = m_o > 0$ in rectangular brackets maintain hidden active - passive gravimechanical partnerships under the strict equalities (7) for active (source) and passive (sink) fractions of the {\it gesamt} GR energy.
The inseparable bound unlike active comasses $m^+_a (+x^o) = - M sin \gamma $ and $m^-_a (-x^o) = -M sin (- \gamma )$ cross generate Maxwell-type fields for attraction of bound passive comasses in the same 3D space but under reverse coordinate rates $\mp dx^o$, respectively. Therefore, (mechanical) passive-inertial  flows are cross-balanced by active EM-type gravitational field flows.   This inverted partnership from dynamical Lagrange identities (11) of folded time reality clarifies negative active masses of radial sources  under positive passive-inertial masses of radial sinks. An inversion of the folded forth coordinate, $+x^4 \rightarrow - x^4$,
changes conventional source (active) and sink (passive) notions in the Lagrangian, but keeps the same
physical time and observable space-time dynamics of inertial matter. Folded time symmetry for the Lagrangian maintains creation of residual (inertial) four-momentums under translations of real unlike comasses and residual (inertial) angular momentums under their  rotations. In fact, FTS pseudo-geometry of the 1908 Minkowski four-interval for any inertial masses justified independent  chiral options for left and right  rotations of matter in all modern theories, including Quantum Chromodynamics.

Again, the observable four-current (10) of collinear and equal mechanical and gravitational mass densities matches Coulomb-type cross-branch interactions of active and passive unlike masses in folded time reality with $MPX^o$ symmetry. An increase of the mechanical four-current  of inertial mass is always accompanied by equal  increase of the gravitational mass four-current.  Einstein's Principle of Equivalence of active and passive masses is well established. This Principle originates from strictly balanced imaginary mass flows in iEM vacuum with regular 4D geometry. Broken iEM vacuum geometry (due to the Big Ban fluctuation into folded pseudo-geometry) transfers zero-sum counter-vectors into equal co-vectors which obey the GR Principle of Equivalence.
 Contrary to four cross-balanced active and passive GR mass co-flows in (11), active and passive iEM mass counter-flows are directly balanced in the EM equalities (5) without the cross-branch cooperation. Therefore, the Lagrange variational balance can be achieved for independent motion of unlike electric charges (imaginary masses) that admits their spatial split (observed in practice). The similar spatial split is impossible for two real comasses, which have accompany together each separated electric charge. Again, the point is that  the GR vector balance (11) requires FTS dynamics with mirror local coupling of negative and positive active/passive mass densities.

 One can imagine a  hydrodynamic thought experiment in order to visualize how two unlike comasses can coexist in the same spatial structure of an observed body or a {\it gesamt} carrier of elementary energies.
 The attraction of sink-ends of hoses pumping water from a tank (one side of the universe) and the repulsion of other source-ends of these hoses pumping this water into another tank (second side of the universe with so far the same time parameter) may shed some light on paired unlike comasses in folded time reality.  If the time parameter would be reversed for the second tank, then identical attractions
 of sink-ends could be simultaneously achieved in both sides of this tank-universe (which  can  be folded due to identical dynamics of folded hoses with equal attractions of their ends).

 Equalities (10) for collinear active and passive energy-momentums are compatible with the inward Newtonian field, which produces negative Gauss flux around the active (negative) mass.
 The folded pseudo-balances (11) can take place in real time world only due to identical cross-branch cancellations of two active and two passive mass four-flows.
 The gravitational two-body attraction can be interpreted as paired FTS Coulomb-type interactions of unlike  active and passive comasses. Spatial split of unlike electric charges still holds paired residual comasses in each electric charge. The electron (or another charged object) carries only one unpaired imaginary mass (or electric charge) but twin unlike real comasses with their inseparable coupling. One can combine the imaginary mass flow equalities (5) for electrodynamics with residual mass flow equalities (11) for real time computations of electromagnetic and gravitational fields of the world current density ensemble of continuous overlapping particles (vector flows $n_ku^\mu$ of charged radial structures and their arbitrary fluctuation $l_k^\mu$),
\begin {eqnarray}
\cases {
\pmatrix {  \sum_{_1}^\infty iq_kn u^\mu    \cr\cr \sum_{_1}^\infty 2|m_k|n u^\mu }^{passive}_{{\bf x}, t}
\equiv
- \frac {c^2}{4\pi G} \nabla_\nu  \pmatrix { \sum_{_1}^\infty  { f_k^{\mu\nu}} (iq_k)^*/|m_k|  \cr\cr
   \sum_{_1}^\infty 2{f_k^{\mu\nu}}}_{{\bf x}, t}^{active} \cr
\cr
 \pmatrix {(ic^2/ G )[\nabla_\lambda (\sum_{_1}^\infty f^k_{\mu\nu}  q_k / |m_k|  ) + \nabla_\nu (\sum_{_1}^\infty f^k_{\lambda\mu} q_k / |m_k|)  + \nabla_\mu (\sum_{_1}^\infty f^k_{\nu\lambda}  q_k /|m_k|)  ] \cr \cr   
 \nabla_\lambda (\sum_{_1}^\infty 2f^k_{\mu\nu}c^2) + \nabla_\nu (\sum_{_1}^\infty 2f^k_{\lambda\mu}c^2) + \nabla_\mu (\sum_{_1}^\infty 2f^k_{\nu\lambda}c^2) }_{{\bf x}, t} \equiv \pmatrix {0 \cr\cr 0} \cr
   \cr
\nabla_\mu  \pmatrix { \sum_{_1}^\infty iq_k n u^\mu  
 \cr\cr  \sum_{_1}^\infty 2|m_k| n u^\mu}_{{\bf x}, t}
\equiv \pmatrix {0 \cr\cr 0}
\cr   
 } \end {eqnarray}
Here $i^*q_k$ are active electric charges, $iq_k$ are passive electric charges, and $|m_k| \equiv |m_k^{\pm}(\pm x^o)|$ are elementary rest mass scalars for active and passive twin comasses, with $|m^{\pm}_k (\pm dx^o) + m^{\mp}_k (\mp dx^o)| = 2|m_k| dt > 0 $ and $k = 1,2, ... \infty $. Two unlike comasses of the elementary mechanical current  cross-generate the doubled real time gravitational field intensity $2f_k^{\mu\nu}({\bf x}, t)c^2$
next to the single EM field intensity
$(q_kc^2/Gm_k)f_k^{\mu\nu}({\bf x}, t)$ $\equiv \Lambda_k f_k^{\mu\nu}({\bf x}, t)$
from one unpaired charge $iq_k$.
All fields and covariant derivatives in the FTS equalities (12) are functions of
real (folded)
 physical time, with $\nabla_{\mu =o} = \nabla_t$ and $\partial/ \partial t = \partial / |\pm \partial x^o|$, rather than one of two pseudo-coordinates  $\pm x^o$. Contrary to the vector pseudo-balance of real mass currents,  imaginary mass currents are balanced in (12) in a line with the geometrical identities (7).  Electric charges do not contribute to the real residual masses $m_k$ or to gravitation and inertia of complex masses with local coupling of real unlike comasses.

The real-time equalities (12) for identical active and passive overlapping mass densities can be interpreted through  net Maxwell and Newton fields,
$F^{\mu\nu}({\bf x}, t) \equiv - \sum_k f_k^{\mu\nu}({\bf x}, t) \Lambda_k $ and $G^{\mu\nu} ({\bf x}, t)\equiv - \sum_k f_k^{\mu\nu}({\bf x}, t)c^2 $,
respectively, verses net  electric and mechanical current densities $J_q^\mu ({\bf x}, t)= \sum_k q_kn u^\mu$
and $J_m^\mu ({\bf x}, t)= \sum_k |m_k| n u^\mu $, respectively. The strong conservation laws
$\nabla_\mu J_q^\mu ({\bf x}, t) \equiv  0$ and $\nabla_\mu J_m^\mu ({\bf x}, t) \equiv  0$ for these world current densities are followed from the equations-equalities (12) in the conventional form of the classical theory of real-time fields,
\begin {eqnarray}
\cases { 4 \pi J_q^\mu ({\bf x}, t) = \nabla_\nu  F^{\nu\mu}({\bf x}, t)  \equiv
\sum_{_1}^\infty  { \nabla_\nu f_k^{\mu\nu}}({\bf x}, t) \Lambda_k  \cr\cr
4 \pi G J_m^\mu ({\bf x}, t) = - \nabla_\nu  G^{\nu\mu}({\bf x}, t)  \equiv
- \sum_{_1}^\infty  {\nabla_\nu f_k^{\mu\nu}({\bf x}, t)}c^2. \cr
 }
\end {eqnarray}

The forming-up potential $W_\mu$ in the dimensionless geometrical field $f_{\mu\nu}$ of every elementary energy carrier $k$ with the radial scale $r_{ok}= Gm_k/c^2$ keeps basic identities (7)  under complex gauge transformations, $W'_\mu \rightarrow W_\mu + (C_1 + iC_2)\partial_\mu \chi$. Such a complex gauge composition of gravitational  and electromagnetic four-potentials, $W_\mu$  and  $iA = i const \times W_\mu $, respectively,  addresses the Lagrange dynamics where complex masses may have mutual transformations of electromagnetic and gravimechanical energies.

Now we look at wave solutions ${\tilde W_\mu} ({\bf x}, x^o)
= C_\mu {\rm exp} (k_\nu x^\nu)$  in (5) or (11) in order to discuss similar solutions
${\tilde W_\mu} ({\bf x}, t)$ 
for FTS physics of real time relations (12)-(13). Elementary waves `propagate' within the distributed source+sink carrier of elementary energy. These waves can modulate, in principle, a steady radial density $n(r) = r_o/4\pi r^2(r + r_o)^2$ of the continuous electron. However, transverse wave modulations do not contribute to (identical) active and passive four-currents, because $\nabla {\tilde f_{tr}^{\mu\nu}} = 0$ for then in (7). Such zero-current waves can exist independently within active (source) field of elementary radial matter by keeping the steady radial structure $n(r)$ of passive (sink) fraction of it. But in general, non-empty space physics of overlapping continuous carriers admits also longitude periodic modulations of active (source)  and passive (sink) carrier's fractions, providing they have equal four-vector currents. Longitude electromagnetic waves modulate the steady radial distribution $n(r)$, while FTS longitude waves of real radial comasses can be unavailable for direct observations. In other words, folded time symmetry of non-local comasses provides more options for wave phenomena in non-empty material space than regular physics of tranverse waves in empty-space. For example, one can infer the particle-wave duality of the distributed elementary carrier of nonlocal energy-matter as well as chiral wave doublets with zero Poynting vector under real time observations. Here two equal pseudo-coordinates restore causality of `advanced' gravimechanical waves in practice.

Recall that a EM field of one electric charge is accompanied by twin GR fields of two unlike comasses. Therefore, one vector electromagnetic wave (the photon) is associated in the real-time equation (12) with two paired vector gravitational waves. One may say that the spin 1 EM wave of the accelerated electron is accompanied by paired vector cowaves (of twin unlike comasses) with net spin 2 in unfolded 4D geometry.
Paired emitted gravitational waves of active comasses at mirror branches $\pm x^o$ cross-interact with twin passive comasses during one elementary interaction process. There are no physical options to split unlike twin comasses and, consequently, there are no physical options in observed reality to split zero energy-momentum and zero spin of two paired vector waves. Such a doublet of vector gravitational waves forms a zero spin tensor graviton (or a null-particle).  A positive momentum of emitted cross-waves in the graviton is accompanied by the net positive momentum of emitting active comasses in agreement with (11). Similarly, inertial comasses gain net negatively directed momentum by cross-absorbing this gravitational wave doublet (or the null-particle). The same observation rule applies to angular momentum exchanges. Namely, two gravitating bodies gain opposite spin 2 elementary impacts from each graviton exchange under an instantaneous emission-absorption process for nonlocal matter.

In the FTS description of single-time reality, the unbroken EM vacuum of balanced imaginary massses has unfolded 4D geometry with equal active (source-particle) and passive (sink-particle) four flows. Real time dynamics for complex GR+iEM masses corresponds to folded pseudo-geometry of real twin unlike masses. Their twin consolidation enables residual inertia and observations of vacuum EM energies.  Wave solutions of  (12)-(13)  possess non-vanishing Poynting flux in  folded time reality  only for unpaired EM modulations of complex continuous masses. The strict local coupling of positive and negative twin comasses corresponds to the strict confinement of their paired vector waves (zero tensor gravitons or null-particles) addressed by these comasses exclusively to one or another gravitational partner.
There are no real-time options to screen gravitons between two gravitating bodies or to borrow addressed energy of confined gravitons by third bodies or by interferometers.
At the same time, FTS cowaves with the cross-balanced Poynting vector cross-interact with twin comasses and provide energy-momentum  and information exchanges between (inert and living) material bodies. Quantitative results of these FTS  exchanges of null-particles can be predicted, computed, and observed in practice as mutual chiral interactions.

Cross-branch gravitational energy-momentum and angular momentum exchanges of the radiating two-body system with four active and four passive comasses can be evaluated through electromagnetic wave solutions with $q^2 \rightarrow 4Gm^2$ for the non-relativistic limit.  There is no outward gravitational energy flow from binary stars, for example, under their evolution toward one elementary radial object with the equilibrium density $r_+/4\pi r^2 (r + r_+)^2$ of the coalesced steady state, where $r_+ = r_{o1} + r_{o2}$. There are no thermo-gravitomechanical energy losses under space-time-energy self-organizations of gravitating bodies through zero-spin gravitons or null-particles. Only EM radiation losses can take place due to local transformation of gravitational waves into heat $Q$ within accelerated thermodynamical bodies.
Due to the real-time equalities (12), one can apply the known\cite {Bat} EM wave solution, $dI/d\Omega = (4Gm^2) \times 2a^4 \omega^6 sin^2 \theta (1+sin^2\theta)^2/\pi c^5 = d{\dot Q}/d\Omega,$ to the angular self-heating (of neutron stars) by  paired vector waves under mutual circular rotation of two equal inertial masses m. The coherent application of this EM wave solution of (12) to confined gravitational waves corresponds to the binary system energy conservation (due to heat power accumulation) and to the period decay approximation, $dP/dt = - (48\pi/5c^5)\times (4\pi Gm/P)^{5/3}$,  for the non-relativistic  binary pulsar\cite {Hul} PSR B1913+16.

 \section {Conclusions}

 The strict EM balance of imaginary mass counter-currents and the residual FTS co-currents of real masses in (12)-(13) mean that the main part of EM ($4\times 10^{42}$eV) and GR (0.5 MeV) energies of the electron, for example,  belongs to unfolded 4D geometry of EM vacuum. Practical utilization of EM vacuum energies is given in single-time reality  through real energy-momentum exchanges between inertial masses. Energy-momentum changes of rest-mass carriers stand in practice behind all  observations and measurements. Real inertial masses with positive GR energies do not deny real Coulomb-Lorentz forces between imaginary masses and do not deny real energies of electric charges.  The point is that  EM interactions can be observed only through  inertial dynamics of residual GR mass-energies. Neither imaginary nor real mass flow densities are measured directly in integral energy-momentum exchanges of inertial carriers with discrete (quantized) amounts of energies.
 The proposed approach to waves as periodic modulations of material densities within continuous overlapping carriers of elementary radial (active and passive) energies can be useful for physical interpretation of the world nonlocality in classical terms. FTS phenomena may shed some light on electron's radiation self-acceleration that was not satisfactorily explained yet.

Static pseudo-metric folding of 4D vacuum masses under the constant cosmological length $T$ of the folded $x^o$ coordinate would hold static 3D distributions for elementary mass-energies. What drives the coordinate fold variations $|\pm dx^o|\equiv dt$ as the running time rate $dt\neq 0$?
  There is no well developed theory for the origin of time and its arrow. We may note in the FTS geometrical approach that the main difference between coordinates $x^i$ and $x^o$ in the 4D action (3)  is the finite length interval $2T$ of the pseudo-coordinate $x^o$.
  It might not be illogical to infer that the folded length T is not a constant parameter for the broken vacuum geometry. The real world energy-matter spontaneously appeared due to the Big Bang Fold of zero-energy vacuum states. And partial breaking of 4D  vacuum geometry due to a finite interval $[-cT_{BB}; cT_{BB}]$ folding can be assigned to any of four initially equal coordinates. Such a geometrical fluctuation resulted in residual rest-mass energy creation in infinite 3D space, rather than in one Big Bang center.
  The fluctuation rest-mass density should return to the initial, zero-balanced vacuum energy state. Therefore, particles' rest-masses should monotonously  decay together with the relaxation decrease of the non-equilibrium coordinate folding, $T = |\pm x^o|\rightarrow 0$. The running age of the Universe, $T_U = T_{BB} - T > 0$ and $dT_U = dt > 0$, is ultimately restricted by the initial Big Bang Fold, $T^{max}_U = T_{BB}$, and inertial matter should disappear together with time at the global unfolding point  $T=0$.

    A nonlinear relaxation, $T\rightarrow 0$, of the global pseudo-geometry folding  can result in accelerated or decelerated physical time for evolution of observed matter toward its disappearance.
  The (nonlinear) cosmological decay of all overlapping GR energy-charges can interpret the Hubble red shift thought the GR red shift for light coming from the more dense material Universe in the early epochs of the global geometrical folding. The so called accelerated repulsion of galaxies by unspecified dark energy could be modeled by  nonlinear relaxation of non-equilibrium pseudo-geometry toward regular 4D geometry of EM vacuum filled by pure imaginary masses.  Many other geometrical options for observed matter can be discussed for nonlocal FTS reality.
 Unless an experiment against sink/source elementary identities (7) and global currents-vs-fields equalities (12)-(13) is  found and justified, residual gravitation of complex masses from folded 4D geometry of EM vacuum  could stay in the unification portfolio next to 5D and higher dimension interpretations of charged matter.

\end {document}